\documentclass[sigconf]{acmart}

\AtBeginDocument{%
  \providecommand\BibTeX{{%
    \normalfont B\kern-0.5em{\scshape i\kern-0.25em b}\kern-0.8em\TeX}}}

\copyrightyear{2021} 
\acmYear{2021} 
\setcopyright{acmcopyright}\acmConference[UbiComp-ISWC '21 Adjunct]{Adjunct Proceedings of the 2021 ACM International Joint Conference on Pervasive and Ubiquitous Computing and Proceedings of the 2021 ACM International Symposium on Wearable Computers}{September 21--26, 2021}{Virtual, USA}
\acmBooktitle{Adjunct Proceedings of the 2021 ACM International Joint Conference on Pervasive and Ubiquitous Computing and Proceedings of the 2021 ACM International Symposium on Wearable Computers (UbiComp-ISWC '21 Adjunct), September 21--26, 2021, Virtual, USA}
\acmPrice{15.00}
\acmDOI{10.1145/3460418.3479338}
\acmISBN{978-1-4503-8461-2/21/09}

\begin{document}

%%
%% The "title" command has an optional parameter,
%% allowing the author to define a "short title" to be used in page headers.
\title{Investigating the Reliability of Self-report Data in the Wild: The Quest for Ground Truth}

%%
%% The "author" command and its associated commands are used to define
%% the authors and their affiliations.
%% Of note is the shared affiliation of the first two authors, and the
%% "authornote" and "authornotemark" commands
%% used to denote shared contribution to the research.
\author{Nan Gao}
\email{nan.gao@rmit.edu.au}
\affiliation{%
  \institution{RMIT University}
  \city{Melbourne}
  \country{Australia}
  \postcode{3000}
}

\author{Mohammad Saiedur Rahaman}
\email{saiedur.rahaman@rmit.edu.au}
\affiliation{%
  \institution{RMIT University}
  \city{Melbourne}
  \country{Australia}
  \postcode{3000}
}

\author{Wei Shao}
\email{wei.shao@rmit.edu.au}
\affiliation{%
  \institution{RMIT University}
  \city{Melbourne}
  \country{Australia}
  \postcode{3000}
}

\author{Flora D. Salim}
\email{flora.salim@rmit.edu.au}
\affiliation{%
  \institution{RMIT University}
  \city{Melbourne}
  \country{Australia}
  \postcode{3000}
}

%%
%% By default, the full list of authors will be used in the page
%% headers. Often, this list is too long, and will overlap
%% other information printed in the page headers. This command allows
%% the author to define a more concise list
%% of authors' names for this purpose.
\renewcommand{\shortauthors}{Gao et al.}

%%
%% The abstract is a short summary of the work to be presented in the
%% article.
\begin{abstract}
  Inferring human mental state (e.g., emotion, depression, engagement) with sensing technology is one of the most valuable challenges in the affective computing area, which has a profound impact in all industries interacting with humans. Self-report is the most common way to quantify how people think, but prone to subjectivity and various responses bias. It is usually used as the ground truth for human mental state prediction. In recent years, many data-driven machine learning models are built based on self-report annotations as the target value. In this research, we investigate the reliability of self-report data in the wild by studying the confidence level of responses and survey completion time. We conduct a case study (i.e., student engagement inference) by recruiting 23 students in a high school setting over a period of 4 weeks. Overall, our participants volunteered 488 self-reported responses and sensing data from smart wristbands. We find that the physiologically measured student engagement and perceived student engagement are not always consistent. The findings from this research have great potential to benefit future studies in predicting engagement, depression, stress, and other emotion-related states in the field of affective computing and sensing technologies.
\end{abstract}

%%
%% The code below is generated by the tool at http://dl.acm.org/ccs.cfm.
%% Please copy and paste the code instead of the example below.
%%


%%
%% Keywords. The author(s) should pick words that accurately describe
%% the work being presented. Separate the keywords with commas.

%% A "teaser" image appears between the author and affiliation
%% information and the body of the document, and typically spans the
%% page.

%%
%% This command processes the author and affiliation and title
%% information and builds the first part of the formatted document.
\maketitle

\section{Introduction}

%With the coronavirus (COVID-19) outbreak, the impacts of the measures necessary to contain its spread were likely to negatively impact mental health \cite{australian}. The sudden loss of employment, social interaction, localised 'lockdowns', as well as the pressure of remote work or schooling have affected the mental health of people all over the world. Brooks et al. \cite{brooks2020rapid} pointed that stress, confusion and anger are commonplace as a result of pandemic and WHO \cite{who2021} indicated that, though some people may not experience long-term concerns, COVID-19 may cause or exacerbate long-term mental illness, including anxiety, depression, post-traumatic stress disorder (PTSD), etc. 
%Introduce the prevalence of affective computing and the common approaches for prediction
%https://www.aihw.gov.au/reports/mental-health-services/mental-health-services-in-australia/report-contents/mental-health-impact-of-covid-19#References

In recent decades, with the advances of wearables and IoT devices, sensing technologies have been increasingly investigated to infer human emotion and mental characteristics, which becomes a hot topic in the Ubicomp community, especially surrounding the prediction of mood \cite{Moodexplorer,morshed2019prediction}, depression \cite{wang2018trackingdepression,xu2019leveragingdepression}, stress \cite{king2019microstress}, engagement \cite{gao2020n,huynh2018engagemon,di2018engagement}, concentration \cite{rahaman2020ambient}, personality traits \cite{personalitysensing2018wang,gao2019predicting} etc. Understanding human emotion and mental state with sensing technologies in real-time can help design intervention strategies to prevent mental health issues among people. 

One of the most commonly used methods for measuring emotion and mental state is to ask participants to respond to self-report surveys (e.g., \cite{ gao2019predicting,di2018engagement,gashi2019using}). An alternative to self-report survey is the Ecological Momentary Assessment (EMA), which is designed to repeatedly collect human responses in real-time in natural settings. In the emotion sensing area, the responses from self-report surveys or EMAs are usually regarded as the measure of \textit{ground truth} \cite{king2019microstress,di2018engagement,gao2020n,wang2014studentlife,Moodexplorer} when building the machine learning (ML) prediction model. They are usually served as the target variables while the features extracted from sensing data are used as the predictor in ML contexts. Then, the predictor is mapped to the target variables through the empirical relationship determined by the data. Moller et al. \cite{moller2013investigating} pointed out researchers should not trust the self-reports blindly, but take into consideration that the responses can be unreliable.

%Imagine you are doing a self-report survey. You think and report your answers carefully at first. After answering several questions, you feel bored and want to finish the survey as soon as possible. Then you quickly read the questions and choose one answer randomly. Once you submit, you feel relaxed, as if you have thrown off a big baggage. 

%Why self-report survey is important and popular?

%Predicting human attitude/behaviour using self-report survey is risky...
%https://www.mercuryds.com/blog/using-self-report-survey-data-to-predict-human-psychology-is-tough

%The performance of ML model based on self-report surveys are usually low, one of the reason is... Common reasons for the inaccuracy of self-reports are as follows: 
%1.
%2.
%3.
%In this paper, we will focus on * part.

%Reliability of self-report survey in the lab vs wild.

%We are the first to study how the 
In this research, we investigate the reliability of self-report data by investigating the patterns of the reported confidence level and survey completion time. Then we focus on the emotion sensing area and use the learning engagement as an example to compare the physiologically measured engagement and perceived engagement. We conduct a field study in a private high school, and 488 self-report responses and wearable data are collected from 23 student participants over 144 classes and 10 courses for 4 weeks. In sum, our contributions are as follows:
\begin{itemize}
    \item For the first time, we investigate the reliability of self-report data by studying the confidence level of self-reported responses. Then we  compare the confidence level of responses with the survey completion time to better understand the reliability of self-report data.
    %\item For the first time, we find the correlation between the physiological signals between the reliability of surveys. 
    %\item To the best of the knowledge, we are the first to compare the perceived student engagement and physiologically measured engagement. We find the perceived engagement are not always consistent with the physiologically measure engagement. Participants with similar physiological patterns may have different perceived engagement annotations while participants with similar annotations may have very different physiological patterns. We also explore how the different factors affect the physiologically measured engagement and perceived engagement.
    
     \item Taking the student learning engagement as an example, we find the perceived student engagement and physiologically measured engagement are not always consistent.%. Specifically, we identify if the perceived engagement are consistent with the physiologically measure engagement. We also explore how the different factors affect the physiologically measured engagement and perceived engagement.
     
    \item We point out the risk of using subjective annotations as the ground truth, and discuss the possibility to use physiological signals as objective measures of student engagement. 
    %To the best of our knowledge, we are the first to address that we should treat the physiological signals (i.e., EDA data) as the objective measures of student engagement since the annotation data may be not reliable. We find that even the students with similar patterns of EDA signals may report very different perceived engagement annotation scores. 
    %\item We compared the perceived and physiologically based engagement, and found they are not always consistent. We discussed the risk of using subjective annotations as the 'ground truth', and explored the possibility and feasibility of using physiological signals as objective measures of student engagement. The findings from this research has great potential to benefit the future studies in predicting engagement, depression, stress, and other emotion related state in the field of affective computing.
    %\item We proposed a model to predict the reliability of self-report responses using sensing data. After intensive experiments on several public dataset, we found that the prediction performance of human behaviours/attitudes are improved after the reliable responses are selected as the ground truth. 
\end{itemize}

%The reminder of the paper is as follows. In Section \ref{sec:relatedwork}, we reviewed the literature in the reliability of self-report surveys and emotion sensing technologies. Section \ref{sec:Data collection} introduces the procedure of data collection and measures of different items. Then in Section \ref{sec:reliability}, we investigate the reliability of self-report survey by understanding the confidence level of subjective responses, and the contrasting perceived engagement and physiologically measured engagement. Section \ref{sec:conclusion} discusses the findings and points out future directions of emotion sensing area.

\section{Related Works}
\label{sec:relatedwork}

\subsection{Inferring Emotion and Mental State with Sensing Technology}

In Ubicomp community, many studies have assessed human emotion and mental characteristics with sensing technologies (e.g., engagement \cite{gao2020n,huynh2018engagemon}, stress \cite{king2019microstress}, mood \cite{morshed2019prediction,wang2014studentlife}, depression \cite{bakker2011what,wang2018trackingdepression}), which provide an attractive alternative to traditional self-report surveys or EMAs.  King et al. \cite{king2019microstress} proposed a passive sensing framework for detecting pregnant mothers in the wild, with the micro-EMA questions as a measurable ground truth for stress. Similarly, Gao et al. \cite{gao2020n} predicted student learning engagement with physiological sensing data, with the adapted In-class Student Engagement Questionnaire (ISEQ) \cite{fuller2018development} as the ground truth of learning engagement. Wang et al. \cite{wang2018trackingdepression} tracked depression dynamics in college students using  mobile and wearable sensing approaches, with PHQ-4 \cite{kroenke2009phq4} and PHQ-8 \cite{kroenke2009phq8} scores as the ground truth of depression. Zhang et al. \cite{Moodexplorer} detected the human compound emotion from smartphone sensing data with the self-report responses as the ground truth of emotions. It has become a common practice to regard the subjective responses (e.g., EMA, self-report survey) as the ground truth, and features extracted from sensing data are fed into the data-driven model for emotion and mental state prediction.

\subsection{Reliability of Self-report Data}
Many researchers worked on designing or adapting psychology questionnaires to achieve higher validity and reliability and mitigate response biases \cite{barclay2002not,jackson2018stepovers,sonderen2013ineffectiveness, clark2019constructing,hudson2020comparing}. Clark et al. \cite{clark2019constructing} reviewed recent literature for psychological scale validation and Huston et al. \cite{hudson2020comparing} compared the reliability of different forms of self-reported life satisfaction.
Moller et al. \cite{moller2013investigating} explored the reliability of self-reporting responses under different conditions. They conducted a six-week self-reporting study on smartphone usage. They found that self-reports cannot provide the full image of user behaviours and participants could significantly overestimate the duration of app usage. Though they showed the inaccuracy of self-reports, they gave suggestions for the design of a self-report study (e.g., set reminders, not overcharge participants ) instead of solutions to evaluating the reliability of self-reports. Moreover, they used the survey questions related to real-world behaviour (e.g., smartphone usage) which is easier to be quantified compared with subjective attitudes. 

Wash et al. \cite{wash2017can} investigated the agreement between self-report and behaviors. They found that security research based on self-reports is unreliable for certain behaviours. Especially, when the behavior involves awareness rather than actions, people are less able to answer the questions accurately. Similar to \cite{moller2013investigating}, they revealed the unreliability of self-reports through comparing with the actual behaviours.  

Different from previous studies, this research has several advantages: (1) we investigate the reliability of self-report data through the subjective confidence level provided by users; (2) we reveal the risks of using self-report responses as the ground truth, especially for emotion sensing in Ubicomp community, by comparing the physiological measured engagement and perceived engagement.   

%\subsection{Measuring Human Behaviour with Sensing Technology}

%Other researchers worked on design better questionnaires with high validity and reliability. 

\begin{comment}
\section{Research Questions}
\begin{itemize}
    \item RQ1: Is there correlation between quality of responses and survey completion time?
    \item RQ2: Can we infer the quality of self-report responses using EDA data?
    \item RQ3: Can we improve the prediction performance by selecting reliable self-report responses?
\end{itemize}
\end{comment}
\section{Data Collection}
\label{sec:Data collection}
\subsection{Case Study}
We collected a dataset \cite{gao2021ingauge,gao2021understanding} (available on Figshare \footnote{In-Gauge and En-Gage datasets: \href{https://doi.org/10.25439/rmt.14578908}{https://doi.org/10.25439/rmt.14578908}}) from a field study in a high school over 4 weeks  . The study has been approved by the Human Research Ethics Committee at RMIT University, which was furthermore approved by the principal of the high school. We have recruited 23 students (15-17 years old, 13 female and 10 male) and 6 teachers (33-62 years old, 4 female and 2 male) in Year 10. After returning the signed consent forms by teachers and students (and their guardians), the participants were asked to complete an online survey recording their demographic information (e.g., age,  gender, class information, etc.). 

Before the data collection, all \textit{Empatica E4} wristbands  were synchronized with the E4 Manager App from the same laptop to make sure the internal clocks are correct. During the data collection, student participants were asked to wear the wristband on the non-dominating hands at school-time. They were reminded by the class representative to complete online questionnaires (EMAs) three times a day at 11:00, 13:25, 15:35 (right after the 2$^{nd}$, 4$^{th}$, 5$^{th}$ class). For teacher participants, they only need to wear the wristband during their classes and complete the EMA right after their class. 

As a token of appreciation, participants were distributed four movie vouchers for 4-week data collection. Participation in this research project was completely voluntary, and participants were free to withdraw from the project at any stage.
\subsection{Measures}
%To test our hypothesis, we were interested in student 3-dimensional engagement, seating location, physical movement and physiological signals in class. 

\subsubsection{Student multi-dimensional engagement} We used the self-report to collect subjective assessments of student engagement. It is the most commonly used method to measure student engagement, because it can clearly reflect subjective perceptions of students. According to previous studies \cite{fredricks2004school,fredricks2012measurement}, other methods such as interviews, teacher ratings and observations are vulnerable to external factors. The student engagement questionnaire includes 5 items \footnote{Specifically, the questions are: (1) I paid attention in class; (2) I pretended to participate in class but actually not; (3) I enjoyed learning new things in class; (4) I felt discouraged when we worked on something; (5) I asked myself questions to make sure I understood the class content. The question 1,3 and 5 assess the behavioural, emotional and cognitive engagement respectively, where item 2 and 4 indicate the
behavioural and emotional disaffection   \cite{fuller2018development,skinner2009motivational}.} related to the emotional, behavioural, and cognitive engagement of the validated In-class Student Engagement Questionnaires (ISEQ) \cite{fuller2018development}, which was proved to be effective for multidimensional engagement measurement compared with the traditional long survey. Similar to previous studies \cite{huynh2018engagemon,gashi2019using}, we slightly adapted the questions to suit high school classes and make it easier for underage students to understand. 
%In addition, we did not adopt the original question ‘the activities really helped my learning of this topic’ for assessing cognitive engagement in \cite{fuller2018development}, because some high school classes do not have in-class activities. Instead, we use the well-accepted question ’I asked myself questions to make sure I understood the class content’ \cite{moore2006children}, which has been proven to be a good reflection of cognitive engagement. 
In the questionnaire, each item is rated with a 5-point Likert scale from ’strongly disagree’ to ’strongly agree’.

\subsubsection{Confidence level} At the end of the self-report EMA, we asked the participants to choose their confidence level for their previous responses: "Please rate your confidence level for your answers in this survey (optional)". Then the participants need to choose their option from the 5-point Likert scales, where 1 = not confident, 2 = slightly confident, 3 = moderately confident, 4 = very confident, 5 = extremely confident. The default option is 3: moderately confident. We make this question optional rather than mandatory to minimize the possibility of users answering questions randomly.

%Imagine you are doing a self-report survey. You think and report your answers carefully at first. After answering several questions, you feel bored and want to finish the survey as soon as possible. Then you quickly read the questions and choose one answer randomly. Once you submit, you feel relaxed, as if you have thrown off a big baggage. 
\begin{comment}
\begin{figure}
    \centering
    \includegraphics[width=0.38\textwidth]{image/EDA.png}
    \caption{Graphical representation of EDA components}
    \label{fig:edaexample}
\end{figure}
\end{comment}
\subsubsection{Physiological signals}

We assessed participants' physiological signals (EDA, PPG, ACC, ST) using the \textit{Empatica E4} wristbands. PPG sensor measures the blood volume pulse (BVP) at 64 Hz, from which the inter-beat interval (IBI) and heart rate variability (HRV) can be derived. ACC sensor records 3-axis acceleration at 32 Hz to capture motion-based activities. The optical thermometer captures peripheral skin temperature (ST) at 4 Hz.
EDA sensor records the constantly fluctuating changes in the electrical properties of the skin at 4 Hz. When the level of sweat increases, the conductivity of the skin increases. For most people, when they experience increased cognitive workload, emotional arousal or physical exertion, the brain will send innervating signals to the skin to increase sweating. Even though they may not feel any sweat on the skin surface, the conductivity increases noticeably.

%EDA complex includes two main components: general tonic components (Skin Conductance Level, SCL) and rapid phasic components (Skin Conductance Response, SCR) resulted from sympathetic neuron activity \cite{guideeda}. The SCLs relate to the slower acting and background characteristics of the EDA signal (overall level, slow declination or climbing over time), reflecting the general sweat glands influenced by the autonomic arousal. The SCRs usually show up as abrupt increases in the conductance of skin, which are usually associated with short-term events and external/internal stimuli. Figure \ref{fig:edaexample} shows the graphical representation of EDA responses to a hypothetical stimulus, including three stages: latency, rise time and recovery time.

\section{Reliability of Self-report Data}
\label{sec:reliability}

%In this section, we first investigate the overview of response reliability, including the distribution of confidence level across all self-report responses and all participants. Then we study the relationship between survey completion time and response reliability.

\begin{figure}
    \centering
    \includegraphics[width=0.6\linewidth]{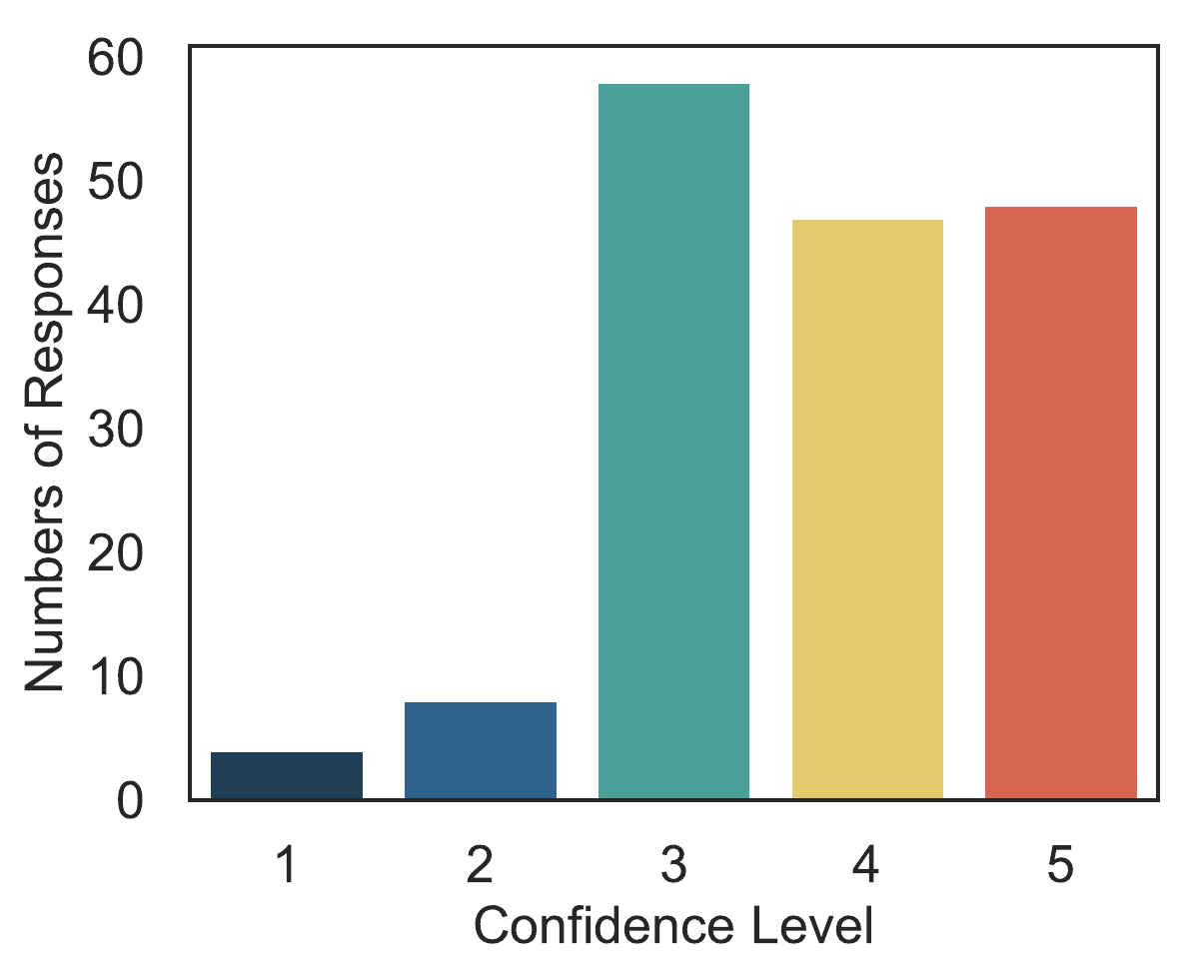}
    \caption{Distribution of confidence level for all self-report responses}
    \label{fig:dis_con}
\end{figure}

\subsection{Confidence Level of Responses}

During the data collection process, we collected the confidence level of self-report from different participants.
Figure \ref{fig:dis_con} shows the distribution of confidence level of different participants. We can see that most participants have a moderate degree of confidence in their responses, but a small number of participants (whose confidence level is 1 or 2) are not very confident in their responses.

\begin{figure}
    \centering
    \includegraphics[width=0.49\textwidth]{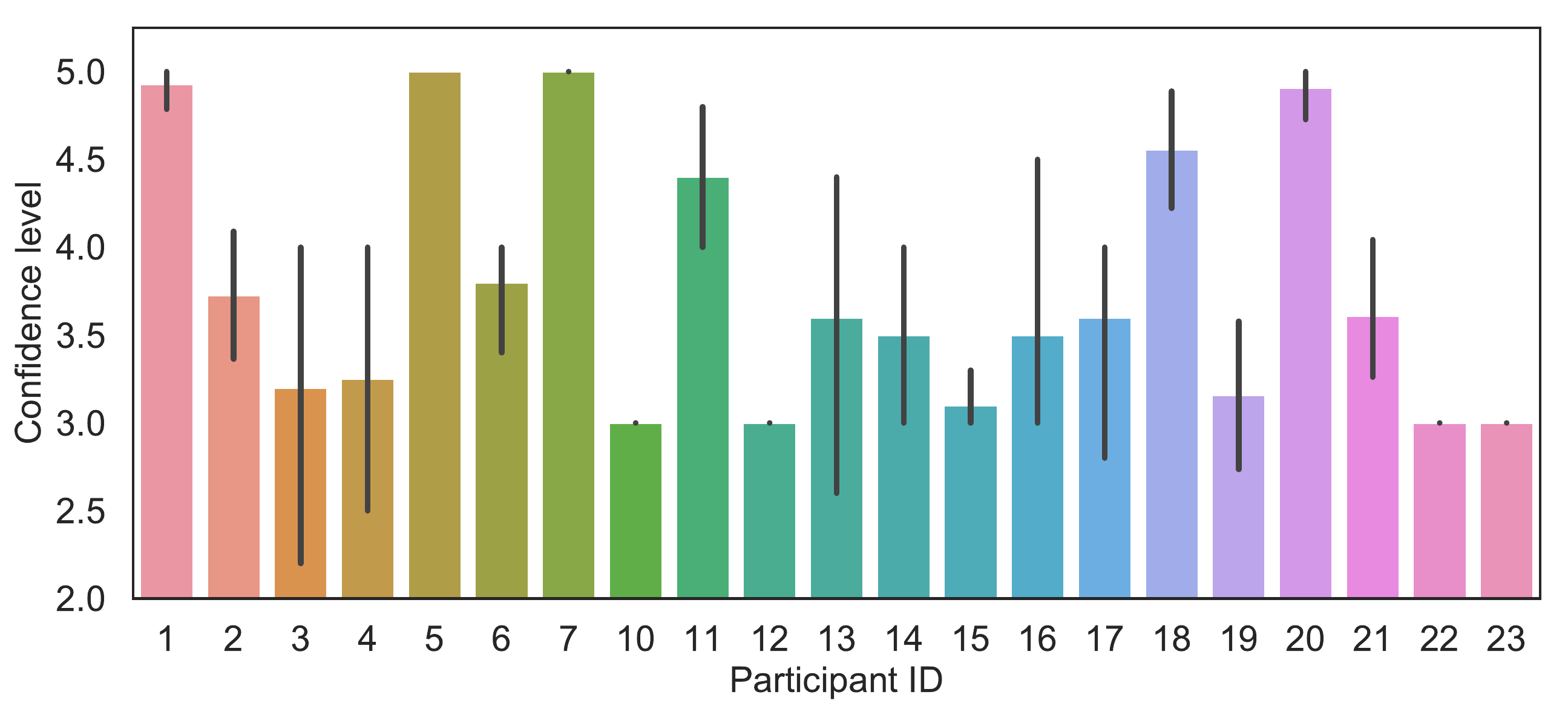}
    \caption{Confidence level across different participants}
    \label{fig:box_confi}
\end{figure}

Then, we investigate whether the same participant tends to have a similar confidence level. Figure \ref{fig:box_confi} shows the boxplot of confidence level across different participants. We find that different participants tend to have very different confidence levels. For example,  some participants (e.g., P1, P20) are usually strongly confident (>4) in their self-report responses, while some participants (e.g., P10, P12, P15) are generally not very confident in their responses. In addition, some participants (e.g., P1, P20, P15) tend to have similar confidence levels in longitudinal studies but some participants (e.g., P16, P3) have very different confidence levels at different times of data collection. The above phenomenon is in line with our daily experience.

\subsection{Completion Time and Reliability}

Malhotra et al. \cite{malhotra2008completion} found that the survey completion time is one of the indicators of response quality, although it is affected by multiple factors and varies from person to person. In this research, for each self-reported questionnaire, we collected the completion time automatically recorded by the \textit{Qualtrics} timing question, which is a hidden question added to the questionnaire to track the time spent by the respondent on that page. 

\begin{figure}
    \centering
    \includegraphics[width=0.49\textwidth]{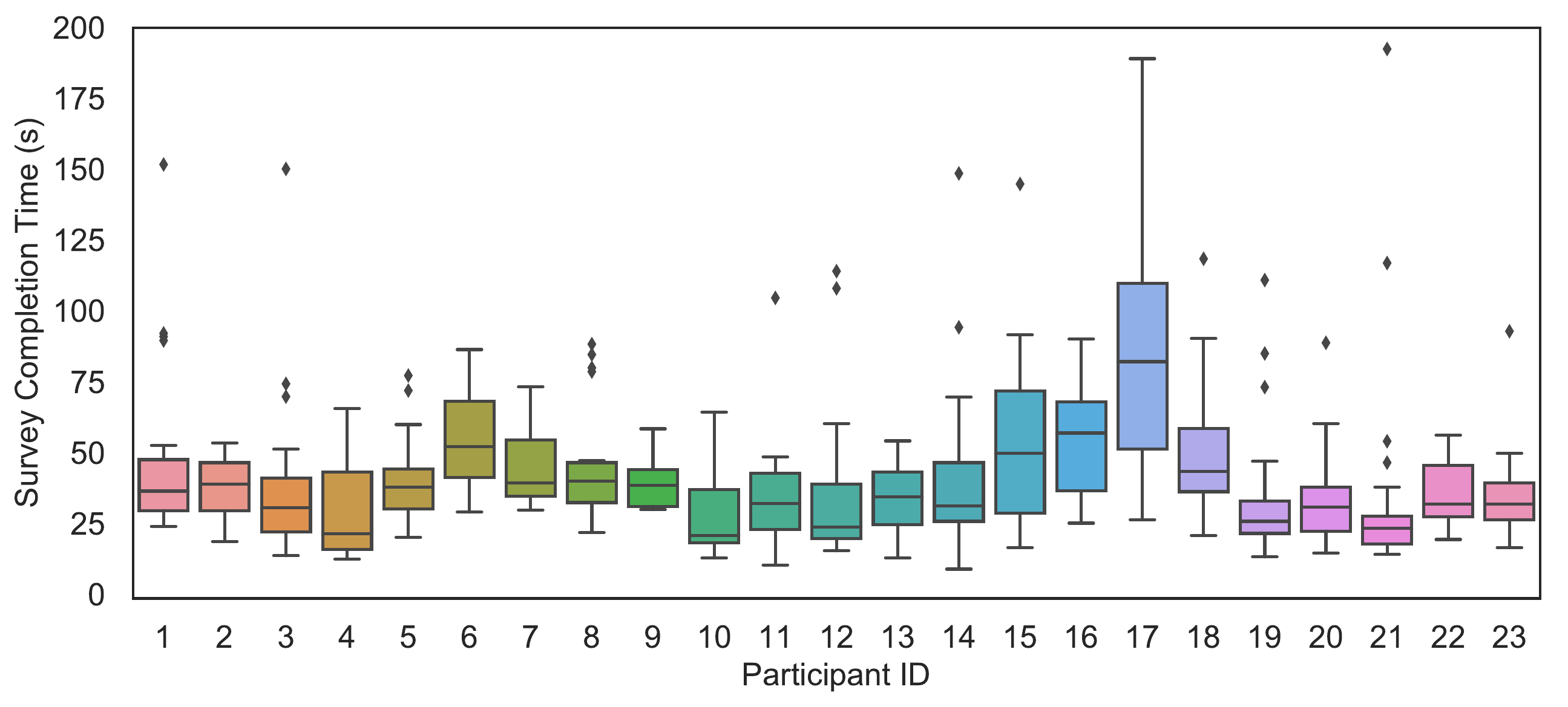}
    \caption{Survey completion time of participants}
    \label{fig:box_time}
\end{figure}

Figure \ref{fig:box_time} shows the survey completion time for all participants. We can see that different participants have very different survey completion time. Most participants complete the survey between 30 to 50 seconds, however, some participants (e.g., P17) spend a lot more time to complete the survey and some participants (e.g., P10, P12) complete the survey in a very short time.

%Most participants complete the survey in 30 to 50 seconds, but some participants complete the survey in less than 15 seconds. Though the survey completion time may be affected by many factors and varies from person to person, it is still one of the indicators of response quality \cite{malhotra2008completion}.

\begin{figure}
    \centering
    \includegraphics[width=0.50\linewidth]{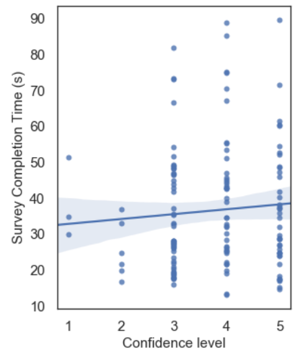}
    \caption{Linear regression of survey completion time with confidence levels }
    \label{fig:regression}
\end{figure}

%In future research, it will be interesting to explore patterns from survey completion time data and assign appropriate weights to survey response for more accurate prediction of student engagement. 
Then we study whether the survey completion time is correlated with the confidence levels. Figure \ref{fig:regression} shows that the survey completion time is positively related to the confidence level. Participants with higher survey completion time tend to have a higher confidence level of the survey. We also investigate how the confidence level correlated with other factors such as the time of the day and weekday, but we do not find the strong correlation between them. In future research, it will be interesting to use survey completion time as an indicator of survey reliability and assign appropriate weights to self-report responses for more accurate human mental-state prediction.
\begin{comment}
\begin{figure}
    \centering
    \includegraphics[width=0.45\linewidth]{image/result.pdf}
    \caption{Prediction performance across different classifiers}
    \label{fig:my_label}
\end{figure}
\end{comment}

%\section{Perceived and physiologically measured engagements}
%\label{sec:contrasting}
%Here, we discuss the risk of using subjective annotations as 'ground truth', using student engagement as the case study. Previously, researchers use the perceived mental state (e.g., emotion, engagement, stress, depression) as ground truth and predicting the perceived state from sensing data \cite{gao2020n,king2019microstress,personalitysensing2018wang}. In this section,  to explore the difference between perceived engagement and physiological measured engagement. 

\subsection{Perceived vs. Physiologically Measured Engagement}

\begin{figure}
    \centering
    \includegraphics[width=0.49\textwidth]{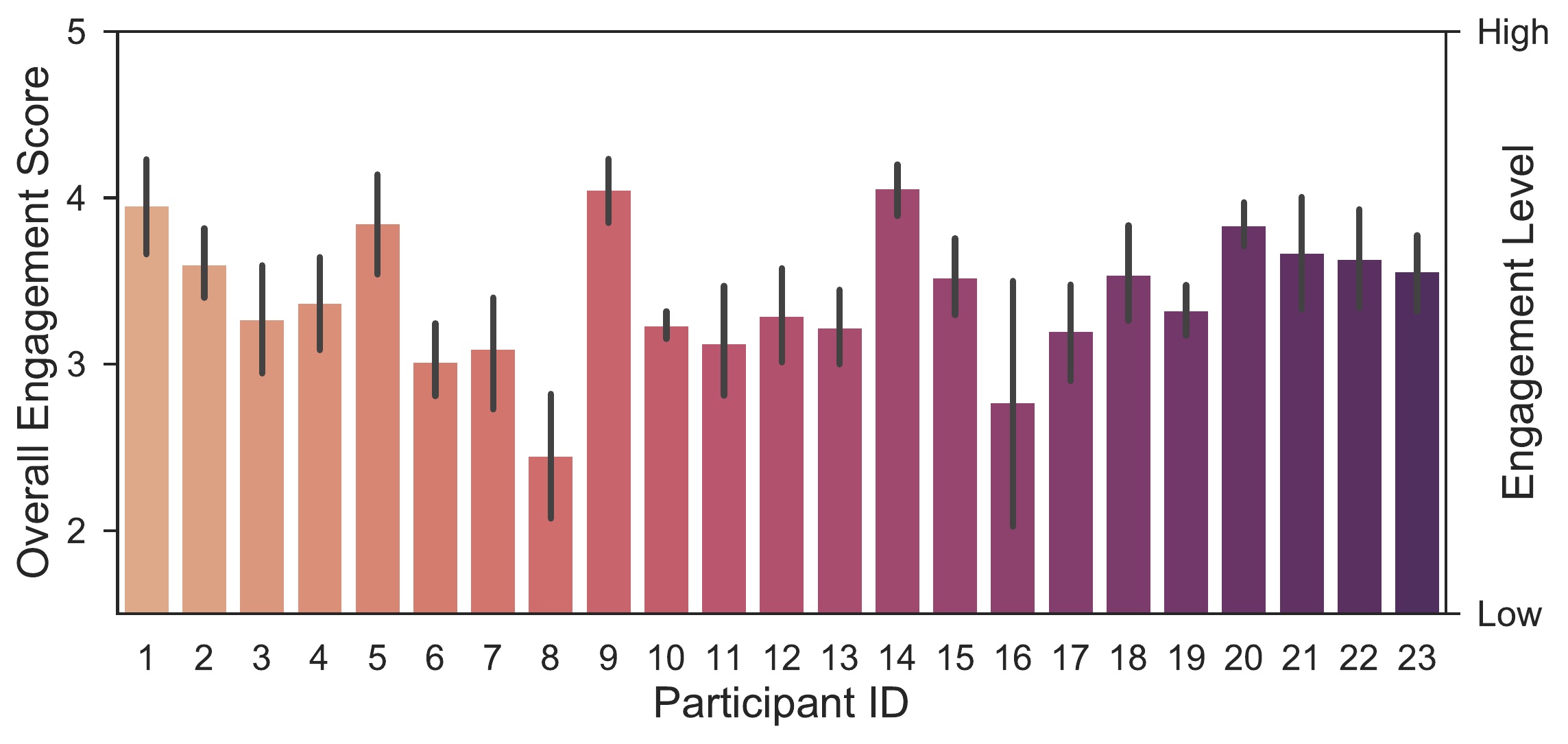}
    \caption{The distribution of overall engagement across student participants}
    \label{fig:dis_engage}
\end{figure}

For the calculation of the perceived engagement scores, we reversed the responses in item 2 and item 4 and then calculated an average score based on the 5-point Likert scale for each dimension of engagement. Then the overall engagement scores were calculated based on all the five items, where 1 indicates the lowest engagement and 5 is the highest engagement. Figure \ref{fig:dis_engage} shows the distribution of overall perceived engagement across student participants. We can see that different participants tend to have very different perceived engagement. Some participants (e.g., P1, P9, P14) are usually highly engaged in the class while some participants (e.g., P8) have low engagement levels. Gao et al. \cite{gao2020n} built the engagement prediction model with the perceived engagement being regarded as the ground truth.

%\textbf{H1:} Different students tend to have different seating preference.

%Figure \ref{fig:loc_dis} displays the seating location across different students participants. We can find that different participants tend to have very different seating preferences. For instance, participant P22 usually seats near the whiteboard while participant P21 tend to seat in the back of the classroom facing the whiteboard.

%\subsection{Physiologically measured engagement}

\begin{figure}
    \centering
    \includegraphics[width=.45\textwidth]{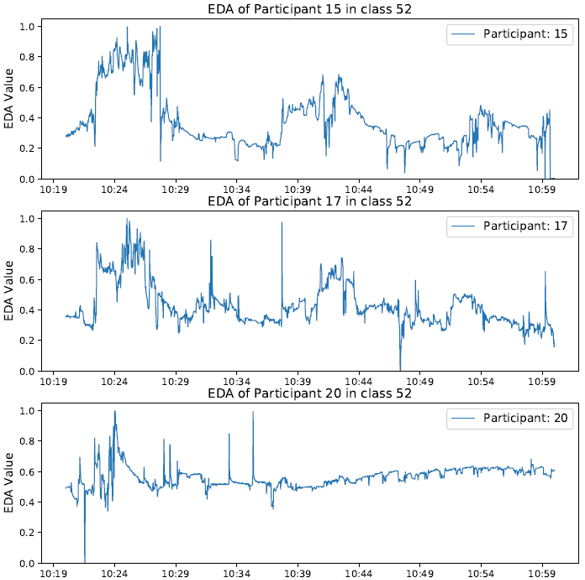}
    \caption{An example of the EDA changes for three different participants P15, P17, P20 in the same class (their perceived engagement are 4.2, 3.2, 4.4)}
    \label{fig:multi_eda}
\end{figure}

\begin{comment}
\begin{figure*}
    \centering
    \includegraphics[width=0.7\textwidth]{image/course_anno.pdf}
    \caption{The impact of course subjects on the perceived self-report engagement}
    \label{fig:impact_subject_anno}
\end{figure*}
\begin{figure*}
    \centering
    \includegraphics[width=0.7\textwidth]{image/course_e4.pdf}
    \caption{The impact of course subjects on the physiological-measured engagement}
    \label{fig:impact_subject}
\end{figure*}

\end{comment}

%\subsubsection{The impact of class length}

%We studied the impact of class length on the perceived engagement and physiological-measured engagement. 

Physiological signals (e.g., EDA, HR, ST signals) have been explored in previous studies to infer student engagement level \cite{gao2020n,di2018engagement}. For example, the EDA level is usually considered  a good indicator of physiological and psychological arousal (e.g., student engagement \cite{gao2020n}, emotional state \cite{di2018engagement}). Increased heart rate indicates the increased efforts and is used as an indirect measure of engagement \cite{richardson2020engagement}. It has been shown that changes in heart rate are related to greater mental efforts and higher information processing demands. Additionally, changes in skin temperature were shown to be correlated with social and mood context \cite{ioannou2014thermal}. 

We show an example of EDA changes for different participants in the same class in Figure \ref{fig:multi_eda}. It can be seen that the EDA signals of the first two participants are very similar and there is a strong physiological synchrony  \cite{palumbo2017interpersonal} between them. Physiological synchrony refers to the association or interdependence of physiological activity between two or more individuals, which has been found in many scenarios. Physiological synchrony between individuals can be indicative of group engagement \cite{palumbo2017interpersonal}, and has been used to measure the classroom emotional climate \cite{gashi2018using} and quantify participants' agreement on self-report engagement \cite{gashi2019using}. 

In Figure \ref{fig:multi_eda}, strong physiological synchrony between P15 and P17 indicates they have similar engagement patterns. Additionally, they both are likely to be highly engaged because (1) their EDA signals have multiple peaks at a similar time, which is a good indicator of emotion arousal; (2) if they are not engaged in class, their EDA changes should be more random instead of being similar. What's more, participant P20 is likely to have lower engagement than participants P17 and P20 since the EDA signal of P20 is more random and the number of peaks is not as many as that of the other two participants. However, based on the self-report responses, the engagement score of three participants P15, P17 and P20 are 4.2, 3.2 and 4.4. From this example, we find that (1) participants with very similar physiological patterns may have very different perceived engagement annotations (see P15 and P17); (2) participants with very similar annotations may have very different physiological patterns (see P15 and P20).

\section{Discussion and Conclusion}
\label{sec:conclusion}
Self-report is one of the most common ways to study the human psychological state and attitude in human-based studies. In the affective computing area, self-report annotations are usually served as the ground truth for predicting human mental state with sensing technologies. Especially in recent years, various data-driven models are built with self-report data as the target variable. However, self-report data is prone to  subjectivity and various responses bias, making it risky and inaccurate to be used as the ground truth in predicting the psychological state (e.g., emotion, depression, engagement, etc.) from sensing data.

In this research, we investigate the reliability of self-report data in the wild from two aspects: (1) For the first time, we study the confidence level of self-report responses from participants, and compare the confidence level with the survey
completion time to better understand the reliability of self-report data; (2) To the best of our knowledge, we are the first to compare the perceived and physiologically measures of student engagement. We find that the perceived self-report engagement are not always consistent with the physiologically measured engagement. Participants with similar physiological patterns may report very different perceived engagement and participants with similar self-report annotations may also have very different physiological patterns. By contrasting the self-report and physiological measures, we reveal the potential risks of only using subjective annotations as the ground truth. %The findings from this research have great potential to benefit future research in the mental health and emotion sensing area. 

%Different factors are explored to study how they affect the physiologically measured and perceived engagement at the same time.

%To the best of our knowledge, we are the first to address that we should treat the physiological signals (i.e., EDA data) as the objective measures of student engagement.

This research is a very promising step towards the study of reliability of self-report data in the wild. It serves as a wake-up call for the emotion and mental sensing research in the Ubicomp community which usually regards the self-report annotations as the ground truth for predicting human mental state. Why do students feel more engaged in class if their bodies say otherwise? Should we trust their subjective self-report responses more, or their objective physiological responses? Is there a better way to understand and model human mental state instead of only using self-report annotations as the ground truth? We hope that more research will be done to explore this issue in the future.

%\section{Acknowledgments}

%This research is supported by the Australian Government through the Australian Research Council's Linkage Projects funding scheme (project LP150100246).

\begin{acks}
This research is supported by the Australian Government through the Australian Research Council's Linkage Projects funding scheme (project LP150100246).
\end{acks}

%%
%% The next two lines define the bibliography style to be used, and
%% the bibliography file.
\bibliographystyle{ACM-Reference-Format}
\bibliography{Nan.bib}

%%% -*-BibTeX-*-
%%% Do NOT edit. File created by BibTeX with style
%%% ACM-Reference-Format-Journals [18-Jan-2012].

\begin{thebibliography}{34}

%%% ====================================================================
%%% NOTE TO THE USER: you can override these defaults by providing
%%% customized versions of any of these macros before the \bibliography
%%% command.  Each of them MUST provide its own final punctuation,
%%% except for \shownote{}, \showDOI{}, and \showURL{}.  The latter two
%%% do not use final punctuation, in order to avoid confusing it with
%%% the Web address.
%%%
%%% To suppress output of a particular field, define its macro to expand
%%% to an empty string, or better, \unskip, like this:
%%%
%%% \newcommand{\showDOI}[1]{\unskip}   % LaTeX syntax
%%%
%%% \def \showDOI #1{\unskip}           % plain TeX syntax
%%%
%%% ====================================================================

\ifx \showCODEN    \undefined \def \showCODEN     #1{\unskip}     \fi
\ifx \showDOI      \undefined \def \showDOI       #1{#1}\fi
\ifx \showISBNx    \undefined \def \showISBNx     #1{\unskip}     \fi
\ifx \showISBNxiii \undefined \def \showISBNxiii  #1{\unskip}     \fi
\ifx \showISSN     \undefined \def \showISSN      #1{\unskip}     \fi
\ifx \showLCCN     \undefined \def \showLCCN      #1{\unskip}     \fi
\ifx \shownote     \undefined \def \shownote      #1{#1}          \fi
\ifx \showarticletitle \undefined \def \showarticletitle #1{#1}   \fi
\ifx \showURL      \undefined \def \showURL       {\relax}        \fi
% The following commands are used for tagged output and should be
% invisible to TeX
\providecommand\bibfield[2]{#2}
\providecommand\bibinfo[2]{#2}
\providecommand\natexlab[1]{#1}
\providecommand\showeprint[2][]{arXiv:#2}

\bibitem[\protect\citeauthoryear{{Bakker}, {Pechenizkiy}, and
  {Sidorova}}{{Bakker} et~al\mbox{.}}{2011}]%
        {bakker2011what}
\bibfield{author}{\bibinfo{person}{Jorn {Bakker}}, \bibinfo{person}{Mykola
  {Pechenizkiy}}, {and} \bibinfo{person}{Natalia {Sidorova}}.}
  \bibinfo{year}{2011}\natexlab{}.
\newblock \showarticletitle{What's Your Current Stress Level? Detection of
  Stress Patterns from GSR Sensor Data}. In \bibinfo{booktitle}{\emph{2011 IEEE
  11th International Conference on Data Mining Workshops}}.
  \bibinfo{pages}{573--580}.
\newblock


\bibitem[\protect\citeauthoryear{{Barclay}, {Todd}, {Finlay}, {Grande}, and
  {Wyatt}}{{Barclay} et~al\mbox{.}}{2002}]%
        {barclay2002not}
\bibfield{author}{\bibinfo{person}{Stephen {Barclay}}, \bibinfo{person}{Chris
  {Todd}}, \bibinfo{person}{Ilora {Finlay}}, \bibinfo{person}{Gunn {Grande}},
  {and} \bibinfo{person}{Penny {Wyatt}}.} \bibinfo{year}{2002}\natexlab{}.
\newblock \showarticletitle{Not Another Questionnaire! Maximizing the Response
  Rate, Predicting Non-response and Assessing Non-response Bias in Postal
  Questionnaire Studies of GPs}.
\newblock \bibinfo{journal}{\emph{Family Practice}} \bibinfo{volume}{19},
  \bibinfo{number}{1} (\bibinfo{year}{2002}), \bibinfo{pages}{105--111}.
\newblock


\bibitem[\protect\citeauthoryear{{Clark} and {Watson}}{{Clark} and
  {Watson}}{2019}]%
        {clark2019constructing}
\bibfield{author}{\bibinfo{person}{Lee~Anna {Clark}} {and}
  \bibinfo{person}{David {Watson}}.} \bibinfo{year}{2019}\natexlab{}.
\newblock \showarticletitle{Constructing validity: New developments in creating
  objective measuring instruments.}
\newblock \bibinfo{journal}{\emph{Psychological Assessment}}
  \bibinfo{volume}{31}, \bibinfo{number}{12} (\bibinfo{year}{2019}),
  \bibinfo{pages}{1412--1427}.
\newblock


\bibitem[\protect\citeauthoryear{Di~Lascio, Gashi, and Santini}{Di~Lascio
  et~al\mbox{.}}{2018}]%
        {di2018engagement}
\bibfield{author}{\bibinfo{person}{Elena Di~Lascio}, \bibinfo{person}{Shkurta
  Gashi}, {and} \bibinfo{person}{Silvia Santini}.}
  \bibinfo{year}{2018}\natexlab{}.
\newblock \showarticletitle{Unobtrusive Assessment of Students' Emotional
  Engagement during Lectures using Electrodermal Activity Sensors}.
\newblock \bibinfo{journal}{\emph{Proceedings of the ACM on Interactive,
  Mobile, Wearable and Ubiquitous Technologies}} \bibinfo{volume}{2},
  \bibinfo{number}{3} (\bibinfo{year}{2018}), \bibinfo{pages}{1--21}.
\newblock


\bibitem[\protect\citeauthoryear{Fredricks, Blumenfeld, and Paris}{Fredricks
  et~al\mbox{.}}{2004}]%
        {fredricks2004school}
\bibfield{author}{\bibinfo{person}{Jennifer~A Fredricks},
  \bibinfo{person}{Phyllis~C Blumenfeld}, {and} \bibinfo{person}{Alison~H
  Paris}.} \bibinfo{year}{2004}\natexlab{}.
\newblock \showarticletitle{School Engagement: Potential of the Concept, State
  of the Evidence}.
\newblock \bibinfo{journal}{\emph{Review of Educational Research}}
  \bibinfo{volume}{74}, \bibinfo{number}{1} (\bibinfo{year}{2004}),
  \bibinfo{pages}{59--109}.
\newblock


\bibitem[\protect\citeauthoryear{Fredricks and McColskey}{Fredricks and
  McColskey}{2012}]%
        {fredricks2012measurement}
\bibfield{author}{\bibinfo{person}{Jennifer~A Fredricks} {and}
  \bibinfo{person}{Wendy McColskey}.} \bibinfo{year}{2012}\natexlab{}.
\newblock \showarticletitle{The Measurement of Student Engagement: A
  Comparative Analysis of Various Methods and Student Self-report Instruments}.
\newblock In \bibinfo{booktitle}{\emph{Handbook of Research on Student
  Engagement}}. \bibinfo{publisher}{Springer}, \bibinfo{pages}{763--782}.
\newblock


\bibitem[\protect\citeauthoryear{Fuller, Karunaratne, Naidu, Exintaris, Short,
  Wolcott, Singleton, and White}{Fuller et~al\mbox{.}}{2018}]%
        {fuller2018development}
\bibfield{author}{\bibinfo{person}{Kathryn~A Fuller},
  \bibinfo{person}{Nilushi~S Karunaratne}, \bibinfo{person}{Som Naidu},
  \bibinfo{person}{Betty Exintaris}, \bibinfo{person}{Jennifer~L Short},
  \bibinfo{person}{Michael~D Wolcott}, \bibinfo{person}{Scott Singleton}, {and}
  \bibinfo{person}{Paul~J White}.} \bibinfo{year}{2018}\natexlab{}.
\newblock \showarticletitle{Development of a Self-report Instrument for
  Measuring in-class Student Engagement Reveals that Pretending to Engage is a
  Significant Unrecognized Problem}.
\newblock \bibinfo{journal}{\emph{PLoS ONE}} \bibinfo{volume}{13},
  \bibinfo{number}{10} (\bibinfo{year}{2018}), \bibinfo{pages}{e0205828}.
\newblock


\bibitem[\protect\citeauthoryear{Gao, Marschall, Burry, Watkins, and Salim}{Gao
  et~al\mbox{.}}{2021a}]%
        {gao2021ingauge}
\bibfield{author}{\bibinfo{person}{Nan Gao}, \bibinfo{person}{Max Marschall},
  \bibinfo{person}{Jane Burry}, \bibinfo{person}{Simon Watkins}, {and}
  \bibinfo{person}{Flora Salim}.} \bibinfo{year}{2021}\natexlab{a}.
\newblock \bibinfo{title}{In-Gauge and En-Gage Datasets}.
\newblock \bibinfo{howpublished}{\emph{Figshare}.
  \url{https://doi.org/10.25439/rmt.14578908}}.
\newblock
\urldef\tempurl%
\url{https://doi.org/10.25439/rmt.14578908}
\showDOI{\tempurl}


\bibitem[\protect\citeauthoryear{Gao, Marschall, Burry, Watkins, and Salim}{Gao
  et~al\mbox{.}}{2021b}]%
        {gao2021understanding}
\bibfield{author}{\bibinfo{person}{Nan Gao}, \bibinfo{person}{Max Marschall},
  \bibinfo{person}{Jane Burry}, \bibinfo{person}{Simon Watkins}, {and}
  \bibinfo{person}{Flora~D. Salim}.} \bibinfo{year}{2021}\natexlab{b}.
\newblock \bibinfo{title}{Understanding Occupants' Behaviour, Engagement,
  Emotion, and Comfort Indoors with Heterogeneous Sensors and Wearables}.
\newblock
\newblock
\showeprint[arxiv]{2105.06637}~[cs.HC]


\bibitem[\protect\citeauthoryear{Gao, Shao, Rahaman, and Salim}{Gao
  et~al\mbox{.}}{2020}]%
        {gao2020n}
\bibfield{author}{\bibinfo{person}{Nan Gao}, \bibinfo{person}{Wei Shao},
  \bibinfo{person}{Mohammad~Saiedur Rahaman}, {and} \bibinfo{person}{Flora~D
  Salim}.} \bibinfo{year}{2020}\natexlab{}.
\newblock \showarticletitle{n-Gage: Predicting in-class Emotional, Behavioural
  and Cognitive Engagement in the Wild}.
\newblock \bibinfo{journal}{\emph{Proceedings of the ACM on Interactive,
  Mobile, Wearable and Ubiquitous Technologies}} \bibinfo{volume}{4},
  \bibinfo{number}{3} (\bibinfo{year}{2020}), \bibinfo{pages}{1--26}.
\newblock


\bibitem[\protect\citeauthoryear{Gao, Shao, and Salim}{Gao
  et~al\mbox{.}}{2019}]%
        {gao2019predicting}
\bibfield{author}{\bibinfo{person}{Nan Gao}, \bibinfo{person}{Wei Shao}, {and}
  \bibinfo{person}{Flora~D Salim}.} \bibinfo{year}{2019}\natexlab{}.
\newblock \showarticletitle{Predicting Personality Traits from Physical
  Activity Intensity}.
\newblock \bibinfo{journal}{\emph{Computer}} \bibinfo{volume}{52},
  \bibinfo{number}{7} (\bibinfo{year}{2019}), \bibinfo{pages}{47--56}.
\newblock


\bibitem[\protect\citeauthoryear{Gashi, Di~Lascio, and Santini}{Gashi
  et~al\mbox{.}}{2018}]%
        {gashi2018using}
\bibfield{author}{\bibinfo{person}{Shkurta Gashi}, \bibinfo{person}{Elena
  Di~Lascio}, {and} \bibinfo{person}{Silvia Santini}.}
  \bibinfo{year}{2018}\natexlab{}.
\newblock \showarticletitle{Using Students' Physiological Synchrony to Quantify
  the Classroom Emotional Climate}. In \bibinfo{booktitle}{\emph{Proceedings of
  the 2018 ACM International Joint Conference and 2018 International Symposium
  on Pervasive and Ubiquitous Computing and Wearable Computers}}.
  \bibinfo{pages}{698--701}.
\newblock


\bibitem[\protect\citeauthoryear{Gashi, Di~Lascio, and Santini}{Gashi
  et~al\mbox{.}}{2019}]%
        {gashi2019using}
\bibfield{author}{\bibinfo{person}{Shkurta Gashi}, \bibinfo{person}{Elena
  Di~Lascio}, {and} \bibinfo{person}{Silvia Santini}.}
  \bibinfo{year}{2019}\natexlab{}.
\newblock \showarticletitle{Using Unobtrusive Wearable Sensors to Measure the
  Physiological Synchrony between Presenters and Audience Members}.
\newblock \bibinfo{journal}{\emph{Proceedings of the ACM on Interactive,
  Mobile, Wearable and Ubiquitous Technologies}} \bibinfo{volume}{3},
  \bibinfo{number}{1} (\bibinfo{year}{2019}), \bibinfo{pages}{1--19}.
\newblock


\bibitem[\protect\citeauthoryear{Hudson, Anusic, Lucas, and Donnellan}{Hudson
  et~al\mbox{.}}{2020}]%
        {hudson2020comparing}
\bibfield{author}{\bibinfo{person}{Nathan~W. Hudson}, \bibinfo{person}{Ivana
  Anusic}, \bibinfo{person}{Richard~E. Lucas}, {and} \bibinfo{person}{M.~Brent
  Donnellan}.} \bibinfo{year}{2020}\natexlab{}.
\newblock \showarticletitle{Comparing the Reliability and Validity of Global
  Self-Report Measures of Subjective Well-Being With Experiential Day
  Reconstruction Measures}.
\newblock \bibinfo{journal}{\emph{Assessment}} \bibinfo{volume}{27},
  \bibinfo{number}{1} (\bibinfo{year}{2020}), \bibinfo{pages}{102--116}.
\newblock
\urldef\tempurl%
\url{https://doi.org/10.1177/1073191117744660}
\showDOI{\tempurl}
\showeprint{https://doi.org/10.1177/1073191117744660}
\newblock
\shownote{PMID: 29254354.}


\bibitem[\protect\citeauthoryear{Huynh, Kim, Ko, Balan, and Lee}{Huynh
  et~al\mbox{.}}{2018}]%
        {huynh2018engagemon}
\bibfield{author}{\bibinfo{person}{Sinh Huynh}, \bibinfo{person}{Seungmin Kim},
  \bibinfo{person}{JeongGil Ko}, \bibinfo{person}{Rajesh~Krishna Balan}, {and}
  \bibinfo{person}{Youngki Lee}.} \bibinfo{year}{2018}\natexlab{}.
\newblock \showarticletitle{EngageMon: Multi-Modal Engagement Sensing for
  Mobile Games}.
\newblock \bibinfo{journal}{\emph{Proceedings of the ACM on Interactive,
  Mobile, Wearable and Ubiquitous Technologies}} \bibinfo{volume}{2},
  \bibinfo{number}{1} (\bibinfo{year}{2018}), \bibinfo{pages}{1--27}.
\newblock


\bibitem[\protect\citeauthoryear{{Ioannou}, {Gallese}, and {Merla}}{{Ioannou}
  et~al\mbox{.}}{2014}]%
        {ioannou2014thermal}
\bibfield{author}{\bibinfo{person}{Stephanos {Ioannou}},
  \bibinfo{person}{Vittorio {Gallese}}, {and} \bibinfo{person}{Arcangelo
  {Merla}}.} \bibinfo{year}{2014}\natexlab{}.
\newblock \showarticletitle{Thermal Infrared Imaging in Psychophysiology:
  Potentialities and Limits}.
\newblock \bibinfo{journal}{\emph{Psychophysiology}} \bibinfo{volume}{51},
  \bibinfo{number}{10} (\bibinfo{year}{2014}), \bibinfo{pages}{951--963}.
\newblock


\bibitem[\protect\citeauthoryear{{Jackson}, {Barton}, {Ashford}, and
  {Abernethy}}{{Jackson} et~al\mbox{.}}{2018}]%
        {jackson2018stepovers}
\bibfield{author}{\bibinfo{person}{Robin~C. {Jackson}}, \bibinfo{person}{Hayley
  {Barton}}, \bibinfo{person}{Kelly~J. {Ashford}}, {and} \bibinfo{person}{Bruce
  {Abernethy}}.} \bibinfo{year}{2018}\natexlab{}.
\newblock \showarticletitle{Stepovers and Signal Detection: Response
  Sensitivity and Bias in the Differentiation of Genuine and Deceptive Football
  Actions.}
\newblock \bibinfo{journal}{\emph{Frontiers in Psychology}}
  \bibinfo{volume}{9} (\bibinfo{year}{2018}), \bibinfo{pages}{2043}.
\newblock


\bibitem[\protect\citeauthoryear{King, Moskowitz, Egilmez, Zhang, Zhang, Bass,
  Rogers, Ghaffari, Wakschlag, and Alshurafa}{King et~al\mbox{.}}{2019}]%
        {king2019microstress}
\bibfield{author}{\bibinfo{person}{Zachary~D King}, \bibinfo{person}{Judith
  Moskowitz}, \bibinfo{person}{Begum Egilmez}, \bibinfo{person}{Shibo Zhang},
  \bibinfo{person}{Lida Zhang}, \bibinfo{person}{Michael Bass},
  \bibinfo{person}{John Rogers}, \bibinfo{person}{Roozbeh Ghaffari},
  \bibinfo{person}{Laurie Wakschlag}, {and} \bibinfo{person}{Nabil Alshurafa}.}
  \bibinfo{year}{2019}\natexlab{}.
\newblock \showarticletitle{Micro-stress EMA: A Passive Sensing Framework for
  Detecting in-the-wild Stress in Pregnant Mothers}.
\newblock \bibinfo{journal}{\emph{Proceedings of the ACM on Interactive,
  Mobile, Wearable and Ubiquitous Technologies}} \bibinfo{volume}{3},
  \bibinfo{number}{3} (\bibinfo{year}{2019}), \bibinfo{pages}{1--22}.
\newblock


\bibitem[\protect\citeauthoryear{{Kroenke}, {Spitzer}, {Williams}, and
  {Löwe}}{{Kroenke} et~al\mbox{.}}{2009a}]%
        {kroenke2009phq4}
\bibfield{author}{\bibinfo{person}{Kurt {Kroenke}}, \bibinfo{person}{Robert~L.
  {Spitzer}}, \bibinfo{person}{Janet~B.W. {Williams}}, {and}
  \bibinfo{person}{Bernd {Löwe}}.} \bibinfo{year}{2009}\natexlab{a}.
\newblock \showarticletitle{An Ultra-brief Screening Scale for Anxiety and
  Depression: The PHQ-4}.
\newblock \bibinfo{journal}{\emph{Psychosomatics}} \bibinfo{volume}{50},
  \bibinfo{number}{6} (\bibinfo{year}{2009}), \bibinfo{pages}{613--621}.
\newblock


\bibitem[\protect\citeauthoryear{{Kroenke}, {Strine}, {Spitzer}, {Williams},
  {Berry}, and {Mokdad}}{{Kroenke} et~al\mbox{.}}{2009b}]%
        {kroenke2009phq8}
\bibfield{author}{\bibinfo{person}{Kurt {Kroenke}}, \bibinfo{person}{Tara~W.
  {Strine}}, \bibinfo{person}{Robert~L. {Spitzer}}, \bibinfo{person}{Janet~B.W.
  {Williams}}, \bibinfo{person}{Joyce~T. {Berry}}, {and}
  \bibinfo{person}{Ali~H. {Mokdad}}.} \bibinfo{year}{2009}\natexlab{b}.
\newblock \showarticletitle{The PHQ-8 as a Measure of Current Depression in the
  General Population}.
\newblock \bibinfo{journal}{\emph{Journal of Affective Disorders}}
  \bibinfo{volume}{114}, \bibinfo{number}{1} (\bibinfo{year}{2009}),
  \bibinfo{pages}{163--173}.
\newblock


\bibitem[\protect\citeauthoryear{Malhotra}{Malhotra}{2008}]%
        {malhotra2008completion}
\bibfield{author}{\bibinfo{person}{Neil Malhotra}.}
  \bibinfo{year}{2008}\natexlab{}.
\newblock \showarticletitle{Completion Time and Response Order Effects in Web
  Surveys}.
\newblock \bibinfo{journal}{\emph{Public Opinion Quarterly}}
  \bibinfo{volume}{72}, \bibinfo{number}{5} (\bibinfo{year}{2008}),
  \bibinfo{pages}{914--934}.
\newblock


\bibitem[\protect\citeauthoryear{M{\"o}ller, Kranz, Schmid, Roalter, and
  Diewald}{M{\"o}ller et~al\mbox{.}}{2013}]%
        {moller2013investigating}
\bibfield{author}{\bibinfo{person}{Andreas M{\"o}ller},
  \bibinfo{person}{Matthias Kranz}, \bibinfo{person}{Barbara Schmid},
  \bibinfo{person}{Luis Roalter}, {and} \bibinfo{person}{Stefan Diewald}.}
  \bibinfo{year}{2013}\natexlab{}.
\newblock \showarticletitle{Investigating Self-reporting Behavior in Long-term
  Studies}. In \bibinfo{booktitle}{\emph{Proceedings of the SIGCHI Conference
  on Human Factors in Computing Systems}}. \bibinfo{pages}{2931--2940}.
\newblock


\bibitem[\protect\citeauthoryear{{Morshed}, {Saha}, {Li}, {D'Mello},
  {Choudhury}, {Abowd}, and {Plötz}}{{Morshed} et~al\mbox{.}}{2019}]%
        {morshed2019prediction}
\bibfield{author}{\bibinfo{person}{Mehrab~Bin {Morshed}},
  \bibinfo{person}{Koustuv {Saha}}, \bibinfo{person}{Richard {Li}},
  \bibinfo{person}{Sidney~K. {D'Mello}}, \bibinfo{person}{Munmun~De
  {Choudhury}}, \bibinfo{person}{Gregory~D. {Abowd}}, {and}
  \bibinfo{person}{Thomas {Plötz}}.} \bibinfo{year}{2019}\natexlab{}.
\newblock \showarticletitle{Prediction of Mood Instability with Passive
  Sensing}.
\newblock \bibinfo{journal}{\emph{Proceedings of the ACM on Interactive,
  Mobile, Wearable and Ubiquitous Technologies}} \bibinfo{volume}{3},
  \bibinfo{number}{3} (\bibinfo{year}{2019}), \bibinfo{pages}{1--21}.
\newblock


\bibitem[\protect\citeauthoryear{Palumbo, Marraccini, Weyandt, Wilder-Smith,
  McGee, Liu, and Goodwin}{Palumbo et~al\mbox{.}}{2017}]%
        {palumbo2017interpersonal}
\bibfield{author}{\bibinfo{person}{Richard~V Palumbo},
  \bibinfo{person}{Marisa~E Marraccini}, \bibinfo{person}{Lisa~L Weyandt},
  \bibinfo{person}{Oliver Wilder-Smith}, \bibinfo{person}{Heather~A McGee},
  \bibinfo{person}{Siwei Liu}, {and} \bibinfo{person}{Matthew~S Goodwin}.}
  \bibinfo{year}{2017}\natexlab{}.
\newblock \showarticletitle{Interpersonal Autonomic Physiology: A Systematic
  Review of the Literature}.
\newblock \bibinfo{journal}{\emph{Personality and Social Psychology Review}}
  \bibinfo{volume}{21}, \bibinfo{number}{2} (\bibinfo{year}{2017}),
  \bibinfo{pages}{99--141}.
\newblock


\bibitem[\protect\citeauthoryear{Rahaman, Liono, Ren, Chan, Kudo, Rawling, and
  Salim}{Rahaman et~al\mbox{.}}{2020}]%
        {rahaman2020ambient}
\bibfield{author}{\bibinfo{person}{Mohammad~Saiedur Rahaman},
  \bibinfo{person}{Jonathan Liono}, \bibinfo{person}{Yongli Ren},
  \bibinfo{person}{Jeffrey Chan}, \bibinfo{person}{Shaw Kudo},
  \bibinfo{person}{Tim Rawling}, {and} \bibinfo{person}{Flora~D Salim}.}
  \bibinfo{year}{2020}\natexlab{}.
\newblock \showarticletitle{An Ambient-physical System to Infer Concentration
  in Open-plan Workplace}.
\newblock \bibinfo{journal}{\emph{IEEE Internet of Things Journal}}
  (\bibinfo{year}{2020}), \bibinfo{pages}{1--1}.
\newblock
\urldef\tempurl%
\url{https://doi.org/10.1109/JIOT.2020.2996219}
\showDOI{\tempurl}


\bibitem[\protect\citeauthoryear{Richardson, Griffin, Zaki, Stephenson, Yan,
  Curry, Noble, Hogan, Skipper, and Devlin}{Richardson et~al\mbox{.}}{2020}]%
        {richardson2020engagement}
\bibfield{author}{\bibinfo{person}{Daniel~C Richardson},
  \bibinfo{person}{Nicole~K Griffin}, \bibinfo{person}{Lara Zaki},
  \bibinfo{person}{Auburn Stephenson}, \bibinfo{person}{Jiachen Yan},
  \bibinfo{person}{Thomas Curry}, \bibinfo{person}{Richard Noble},
  \bibinfo{person}{John Hogan}, \bibinfo{person}{Jeremy~I Skipper}, {and}
  \bibinfo{person}{Joseph~T Devlin}.} \bibinfo{year}{2020}\natexlab{}.
\newblock \showarticletitle{Engagement in Video and Audio Narratives:
  Contrasting Self-report and Physiological Measures}.
\newblock \bibinfo{journal}{\emph{Scientific Reports}} \bibinfo{volume}{10},
  \bibinfo{number}{1} (\bibinfo{year}{2020}), \bibinfo{pages}{1--8}.
\newblock


\bibitem[\protect\citeauthoryear{Skinner, Kindermann, and Furrer}{Skinner
  et~al\mbox{.}}{2009}]%
        {skinner2009motivational}
\bibfield{author}{\bibinfo{person}{Ellen~A Skinner}, \bibinfo{person}{Thomas~A
  Kindermann}, {and} \bibinfo{person}{Carrie~J Furrer}.}
  \bibinfo{year}{2009}\natexlab{}.
\newblock \showarticletitle{A Motivational Perspective on Engagement and
  Disaffection: Conceptualization and Assessment of Children's Behavioral and
  Emotional Participation in Academic Activities in the Classroom}.
\newblock \bibinfo{journal}{\emph{Educational and Psychological Measurement}}
  \bibinfo{volume}{69}, \bibinfo{number}{3} (\bibinfo{year}{2009}),
  \bibinfo{pages}{493--525}.
\newblock


\bibitem[\protect\citeauthoryear{van {Sonderen}, {Sanderman}, and {Coyne}}{van
  {Sonderen} et~al\mbox{.}}{2013}]%
        {sonderen2013ineffectiveness}
\bibfield{author}{\bibinfo{person}{Eric van {Sonderen}},
  \bibinfo{person}{Robbert {Sanderman}}, {and} \bibinfo{person}{James~C.
  {Coyne}}.} \bibinfo{year}{2013}\natexlab{}.
\newblock \showarticletitle{Ineffectiveness of Reverse Wording of Questionnaire
  Items: Let’s Learn from Cows in the Rain}.
\newblock \bibinfo{journal}{\emph{PLoS ONE}} \bibinfo{volume}{8},
  \bibinfo{number}{7} (\bibinfo{year}{2013}).
\newblock


\bibitem[\protect\citeauthoryear{Wang, Chen, Chen, Li, Harari, Tignor, Zhou,
  Ben-Zeev, and Campbell}{Wang et~al\mbox{.}}{2014}]%
        {wang2014studentlife}
\bibfield{author}{\bibinfo{person}{Rui Wang}, \bibinfo{person}{Fanglin Chen},
  \bibinfo{person}{Zhenyu Chen}, \bibinfo{person}{Tianxing Li},
  \bibinfo{person}{Gabriella Harari}, \bibinfo{person}{Stefanie Tignor},
  \bibinfo{person}{Xia Zhou}, \bibinfo{person}{Dror Ben-Zeev}, {and}
  \bibinfo{person}{Andrew~T Campbell}.} \bibinfo{year}{2014}\natexlab{}.
\newblock \showarticletitle{StudentLife: Assessing Mental Health, Academic
  Performance and Behavioral Trends of College Students using Smartphones}. In
  \bibinfo{booktitle}{\emph{Proceedings of the 2014 ACM International Joint
  Conference on Pervasive and Ubiquitous Computing}}. \bibinfo{pages}{3--14}.
\newblock


\bibitem[\protect\citeauthoryear{Wang, Wang, DaSilva, Huckins, Kelley,
  Heatherton, and Campbell}{Wang et~al\mbox{.}}{2018b}]%
        {wang2018trackingdepression}
\bibfield{author}{\bibinfo{person}{Rui Wang}, \bibinfo{person}{Weichen Wang},
  \bibinfo{person}{Alex DaSilva}, \bibinfo{person}{Jeremy~F Huckins},
  \bibinfo{person}{William~M Kelley}, \bibinfo{person}{Todd~F Heatherton},
  {and} \bibinfo{person}{Andrew~T Campbell}.} \bibinfo{year}{2018}\natexlab{b}.
\newblock \showarticletitle{Tracking Depression Dynamics in College Students
  using Mobile Phone and Wearable Sensing}.
\newblock \bibinfo{journal}{\emph{Proceedings of the ACM on Interactive,
  Mobile, Wearable and Ubiquitous Technologies}} \bibinfo{volume}{2},
  \bibinfo{number}{1} (\bibinfo{year}{2018}), \bibinfo{pages}{1--26}.
\newblock


\bibitem[\protect\citeauthoryear{Wang, Harari, Wang, M{\"u}ller, Mirjafari,
  Masaba, and Campbell}{Wang et~al\mbox{.}}{2018a}]%
        {personalitysensing2018wang}
\bibfield{author}{\bibinfo{person}{Weichen Wang}, \bibinfo{person}{Gabriella~M
  Harari}, \bibinfo{person}{Rui Wang}, \bibinfo{person}{Sandrine~R M{\"u}ller},
  \bibinfo{person}{Shayan Mirjafari}, \bibinfo{person}{Kizito Masaba}, {and}
  \bibinfo{person}{Andrew~T Campbell}.} \bibinfo{year}{2018}\natexlab{a}.
\newblock \showarticletitle{Sensing Behavioral Change over Time: Using
  Within-person Variability Features from Mobile Sensing to Predict Personality
  Traits}.
\newblock \bibinfo{journal}{\emph{Proceedings of the ACM on Interactive,
  Mobile, Wearable and Ubiquitous Technologies}} \bibinfo{volume}{2},
  \bibinfo{number}{3} (\bibinfo{year}{2018}), \bibinfo{pages}{1--21}.
\newblock


\bibitem[\protect\citeauthoryear{Wash, Rader, and Fennell}{Wash
  et~al\mbox{.}}{2017}]%
        {wash2017can}
\bibfield{author}{\bibinfo{person}{Rick Wash}, \bibinfo{person}{Emilee Rader},
  {and} \bibinfo{person}{Chris Fennell}.} \bibinfo{year}{2017}\natexlab{}.
\newblock \showarticletitle{Can People Self-Report Security Accurately?
  Agreement Between Self-Report and Behavioral Measures}. In
  \bibinfo{booktitle}{\emph{Proceedings of the 2017 CHI Conference on Human
  Factors in Computing Systems}}. \bibinfo{pages}{2228--2232}.
\newblock


\bibitem[\protect\citeauthoryear{Xu, Chikersal, Doryab, Villalba, Dutcher,
  Tumminia, Althoff, Cohen, Creswell, Creswell, et~al\mbox{.}}{Xu
  et~al\mbox{.}}{2019}]%
        {xu2019leveragingdepression}
\bibfield{author}{\bibinfo{person}{Xuhai Xu}, \bibinfo{person}{Prerna
  Chikersal}, \bibinfo{person}{Afsaneh Doryab}, \bibinfo{person}{Daniella~K
  Villalba}, \bibinfo{person}{Janine~M Dutcher}, \bibinfo{person}{Michael~J
  Tumminia}, \bibinfo{person}{Tim Althoff}, \bibinfo{person}{Sheldon Cohen},
  \bibinfo{person}{Kasey~G Creswell}, \bibinfo{person}{J~David Creswell},
  {et~al\mbox{.}}} \bibinfo{year}{2019}\natexlab{}.
\newblock \showarticletitle{Leveraging Routine Behavior and
  Contextually-filtered Features for Depression Detection among College
  Students}.
\newblock \bibinfo{journal}{\emph{Proceedings of the ACM on Interactive,
  Mobile, Wearable and Ubiquitous Technologies}} \bibinfo{volume}{3},
  \bibinfo{number}{3} (\bibinfo{year}{2019}), \bibinfo{pages}{1--33}.
\newblock


\bibitem[\protect\citeauthoryear{Zhang, Li, Chen, and Lu}{Zhang
  et~al\mbox{.}}{2018}]%
        {Moodexplorer}
\bibfield{author}{\bibinfo{person}{Xiao Zhang}, \bibinfo{person}{Wenzhong Li},
  \bibinfo{person}{Xu Chen}, {and} \bibinfo{person}{Sanglu Lu}.}
  \bibinfo{year}{2018}\natexlab{}.
\newblock \showarticletitle{Moodexplorer: Towards Compound Emotion Detection
  via Smartphone Sensing}.
\newblock \bibinfo{journal}{\emph{Proceedings of the ACM on Interactive,
  Mobile, Wearable and Ubiquitous Technologies}} \bibinfo{volume}{1},
  \bibinfo{number}{4} (\bibinfo{year}{2018}), \bibinfo{pages}{1--30}.
\newblock


\end{thebibliography}

%%
%% If your work has an appendix, this is the place to put it.

\end{document}